\documentclass[12pt]{article}

\usepackage[nosort]{cite}
\usepackage[hyperref,bulletsep]{collect}
\usepackage{graphicx}
\usepackage{pst-all}
\usepackage{bbm}
\usepackage{amsmath}
\usepackage{amssymb}
\usepackage{subfigure}
\usepackage{slashed}

\makeatletter \@addtoreset{equation}{section} \makeatother

\makeatletter
\let\old@startsection=\@startsection
\let\oldl@section=\l@section
\renewcommand{\@startsection}[6]{\old@startsection{#1}{#2}{#3}{#4}{#5}{#6\mathversion{bold}}}
\renewcommand{\l@section}[2]{\oldl@section{\mathversion{bold}#1}{#2}}
\makeatother

\makeatletter
\let\old@makecaption=\@makecaption
\def\@makecaption{\small\old@makecaption}
\makeatother

\renewcommand{\thefootnote}{\arabic{footnote}}
\let\oldPhi=\Phi
\let\oldPsi=\Psi
\let\oldGamma=\Gamma
\let\oldDelta=\Delta
\let\oldSigma=\Sigma
\let\oldTheta=\Theta
\let\oldPi=\Pi
\let\oldUpsilon=\Upsilon
\renewcommand{\Phi}{\mathnormal{\oldPhi}}
\renewcommand{\Psi}{\mathnormal{\oldPsi}}
\renewcommand{\Gamma}{\mathnormal{\oldGamma}}
\renewcommand{\Sigma}{\mathnormal{\oldSigma}}
\renewcommand{\Delta}{\mathnormal{\oldDelta}}
\renewcommand{\Theta}{\mathnormal{\oldTheta}}
\renewcommand{\Pi}{\mathnormal{\oldPi}}
\renewcommand{\Upsilon}{\mathnormal{\oldUpsilon}}

\newcommand{\superN}{\mathcal{N}}
\newcommand{\Action}{\mathcal{S}}

\newcommand{\tr}{\mathop{\mathrm{tr}}}

\newcommand{\diag}{\mathop{\mathrm{diag}}}

\newcommand{\trans}{{\scriptscriptstyle\mathrm{T}}}

\newcommand{\Integers}{\mathbbm{Z}}
\newcommand{\Reals}{\mathbbm{R}}
\newcommand{\Complex}{\mathbbm{C}}

\ifx\genfrac\sdflkaj

\else

\fi
\newcommand{\sfrac}[2]{{\textstyle\frac{#1}{#2}}}
\newcommand{\half}{\sfrac{1}{2}}

\newcommand{\quarter}{\sfrac{1}{4}}

\newcommand{\Half}{\frac{1}{2}}

\newcommand{\rep}[1]{{\mathbf{#1}}}
\newcommand{\matr}[2]{\left(\begin{array}{#1}#2\end{array}\right)}

\newcommand{\grp}[1]{\mathrm{#1}}

\newcommand{\grU}{\grp{U}}
\newcommand{\grSU}{\grp{SU}}
\newcommand{\grSO}{\grp{SO}}

\newcommand{\bigbrk}[1]{\bigl(#1\bigr)}
\newcommand{\Bigbrk}[1]{\Bigl(#1\Bigr)}

\newcommand{\bigsbrk}[1]{\bigl[#1\bigr]}
\newcommand{\Bigsbrk}[1]{\Bigl[#1\Bigr]}

\newcommand{\acomm}[2]{\{#1,#2\}}

\newcommand{\lrabs}[1]{\left|#1\right|}

\newcommand{\eval}[1]{#1|}

\newcommand{\nn}{\nonumber}
\newcommand{\nln}{\nonumber\\}
\newcommand{\nl}[1][0pt]{\nonumber\\[#1]&\hspace{-4\arraycolsep}&\mathord{}}

\newcommand{\earel}[1]{\mathrel{}&\hspace{-2\arraycolsep}#1\hspace{-2\arraycolsep}&\mathrel{}}
\newcommand{\eq}{\earel{=}}

\def\[{\begin{equation}}
\def\]{\end{equation}}

\makeatletter
\def\mr@ignsp#1 {\ifx\:#1\@empty\else #1\expandafter\mr@ignsp\fi}%
\newcommand{\multiref}[1]{\begingroup
\xdef\mr@no@sparg{\expandafter\mr@ignsp#1 \: }%
\def\mr@comma{}%
\@for\mr@refs:=\mr@no@sparg\do{\mr@comma\def\mr@comma{,}\ref{\mr@refs}}%
\endgroup}
\makeatother

\newcommand{\hypref}[2]{\ifx\href\asklfhas #2\else\href{#1}{#2}\fi}

\newcommand{\secref}[1]{Sec.~\multiref{#1}}

\newcommand{\appref}[1]{App.~\multiref{#1}}

\newcommand{\figref}[1]{Fig.~\multiref{#1}}
\renewcommand{\eqref}[1]{(\multiref{#1})}

\ifx\href\asklfhas\newcommand{\href}[2]{#2}\fi

\newcommand{\comma}{\quad,\quad}
\newcommand{\unit}{\mathbbm{1}}

\newcommand{\levi}{\epsilon}

\newcommand{\eps}{\varepsilon}

\newcommand{\be}{\begin{eqnarray}}
\newcommand{\ee}{\end{eqnarray}}

\newcommand{\auxD}{\mathsf{D}}
\newcommand{\deriD}{\mathcal{D}}
\newcommand{\ZZ}{\mathcal{Z}}
\newcommand{\WW}{\mathcal{W}}

\newcommand{\VV}{\mathcal{V}}

\newcommand{\ZZZ}{\mathfrak{Z}}
\newcommand{\WWW}{\mathfrak{W}}
\newcommand{\VVV}{\mathfrak{V}}

\newcommand{\superW}{\mathrm{W}}

\begin{document}

\thispagestyle{empty}
\begin{flushright}\footnotesize
\texttt{arXiv:0806.1519}\\
\texttt{PUPT-2271}\\
\texttt{UUITP-10/08}\vspace{10mm}
\end{flushright}

\renewcommand{\thefootnote}{\fnsymbol{footnote}}
\setcounter{footnote}{0}

\begin{center}
{\Large\textbf{\mathversion{bold}
Superconformal Chern-Simons Theories and AdS$_4$/CFT$_3$ Correspondence
}\par}

\vspace{1.5cm}

\textrm{Marcus Benna$^a$, Igor Klebanov$^{a,b}$, Thomas Klose$^{a,b}$ and Mikael Smedb{\"a}ck}$^{a,c}$ \vspace{8mm} \\
\textit{
$^a$Joseph Henry Laboratories and $^b$Princeton Center for Theoretical Science \\
Princeton University, Princeton, NJ 08544, USA 
} \\
\textit{
\\
$^c$Department of Physics and Astronomy, Division of Theoretical Physics \\
Uppsala University,
Box 803, SE-751 08 Uppsala, Sweden 
} \\
\texttt{\\ mbenna,klebanov,tklose,smedback@princeton.edu}

\par\vspace{14mm}

\textbf{Abstract} \vspace{5mm}

\begin{minipage}{14cm}
We discuss the $\superN=2$ superspace formulation of the $\superN=8$ superconformal Bagger-Lambert-Gustavsson theory, and of the $\superN=6$ superconformal Aharony-Bergman-Jafferis-Maldacena $\grU(N)\times\grU(N)$ Chern-Simons theory. In particular, we prove the full $\grSU(4)$ R-symmetry of the ABJM theory. We then consider orbifold projections of this theory that give non-chiral and chiral $(\grU(N)\times \grU(N))^n$ superconformal quiver gauge theories. We argue that these theories are dual to certain $AdS_4 \times S^7/(\Integers_n\times \Integers_{\tilde{k}})$ backgrounds of M-theory. We also study a $\grSU(3)$ invariant mass term in the superpotential that makes the $\superN=8$ theory flow to a $\superN=2$ superconformal gauge theory with a sextic superpotential. We conjecture that this gauge theory is dual to the $\grU(1)_R\times \grSU(3)$ invariant extremum of the $\superN=8$ gauged supergravity, which was discovered by N. Warner 25 years ago and whose uplifting to 11 dimensions was found more recently.
\end{minipage}

\end{center}

\vspace{0.5cm}

\newpage
\setcounter{page}{1}
\renewcommand{\thefootnote}{\arabic{footnote}}
\setcounter{footnote}{0}

\hrule
\tableofcontents
\vspace{8mm}
\hrule
\vspace{4mm}

\section{Introduction}

In the recent literature there has been a lot of excitement about the work of 
Bagger and Lambert \cite{Bagger:2006sk,Bagger:2007jr,Bagger:2007vi}, and
the closely related work of Gustavsson \cite{Gustavsson:2007vu},
 who succeeded in finding a $2+1$ dimensional superconformal
Chern-Simons theory with the maximal $\superN=8$ supersymmetry and manifest $\grSO(8)$ R-symmetry. 
These papers were inspired in part by the ideas of \cite{Basu:2004ed,Schwarz:2004yj}. 
The original motivation was a search for a theory describing 
coincident M2-branes. An interesting clue emerged 
in \cite{Lambert:2008et,Distler:2008mk} where it was shown that, for a specially chosen level
of the Chern-Simons gauge theory, its moduli space coincides with that of
a pair of M2-branes at the $\Reals^8/\Integers_2$ singularity. 
The $\Integers_2$ acts by reflection of all 8 coordinates and therefore
does not spoil the $\grSO(8)$ symmetry. However, initial attempts to match the moduli space
of the Chern-Simons gauge theory for arbitrary quantized level $k$ with that of M2-branes 
led to a number of puzzles \cite{VanRaamsdonk:2008ft,Lambert:2008et,Distler:2008mk}.
Very recently, these puzzles were resolved by a very interesting modification of the 
Bagger-Lambert-Gustavsson (BLG) theory proposed by Aharony, Bergman, Jafferis and
Maldacena (ABJM) \cite{Aharony:2008ug} which, in particular, allows for a generalization to an arbitrary
number of M2-branes. This opens the possibilities for many extensions of this work, some of
which we begin exploring in this paper.

The original BLG theory is a particular example of a Chern-Simons gauge theory with gauge group $\grSO(4)$,
but the Chern-Simons term has a somewhat unconventional form. However, van Raamsdonk \cite{VanRaamsdonk:2008ft}
 rewrote the BLG theory as an $\grSU(2)\times \grSU(2)$ gauge theory
coupled to bifundamental matter, as summarized in \secref{sec:BL-review}.
He found conventional Chern-Simons terms for each of the $\grSU(2)$ gauge fields 
although with opposite signs, as noted already in \cite{Bandres:2008vf}.
A more general class of gauge theories of this type was introduced by Gaiotto and Witten
\cite{Gaiotto:2008sd} following \cite{Gaiotto:2007qi}. In this formulation the opposite signs for the two $\grSU(N)$
Chern-Simons terms are related to the $\grSU(N|N)$ supergroup structure.
Although the GW formulation generally has only $\superN=4$ supersymmetry,
it was recently shown how to enlarge the supersymmetry by adding 
more hypermultiplets \cite{Hosomichi:2008jd,Aharony:2008ug}. 
In particular, the maximally supersymmetric BLG theory emerges in the $\grSU(2)\times
\grSU(2)$ case when the matter consists of two bi-fundamental hypermultiplets.
Furthermore, the brane constructions presented in \cite{Gaiotto:2008sd,Aharony:2008ug}
indicate that the relevant gauge theories are actually $\grU(N)\times \grU(N)$. The presence
of the extra interacting $\grU(1)$ compared to the original BLG formulation is crucial for the
complete M-theory interpretation \cite{Aharony:2008ug}. 

One of our aims is to present the BLG theory using $\superN=2$ superspace formulation in $2+1$ dimensions, which is quite similar to the familiar $\superN=1$ superspace in $3+1$ dimensions. In such a formulation only the $\grU(1)_R$ symmetry is manifest, while the quartic superpotential has an additional $\grSU(4)$ global symmetry. For a specially chosen normalization of the superpotential, the full scalar potential is manifestly $\grSO(8)$ invariant. In \secref{sec:BL-superspace}, where we establish the superspace formulation of the BLG theory, we demonstrate how this happens through a special cancellation 
involving the F and D terms.\footnote{This phenomenon is analogous to what happens when the $\superN=4$
SYM theory in 3+1 dimensions is written in terms of an $\superN=1$ gauge theory
coupled to three chiral superfields. While only the $\grU(1)_R \times \grSU(3)$ symmetry is manifest
in such a formulation, the full $\grSU(4)\sim \grSO(6)$ symmetry is found in the potential as a result 
of a specific cancellation between the F and D terms.} 

In \secref{sec:N-N} we study its generalizations to $\grU(N)\times \grU(N)$ gauge theory found by ABJM
\cite{Aharony:2008ug}. The quartic superpotential of this $2+1$ dimensional
theory has exactly the same form as in the $3+1$ dimensional theory on $N$ D3-branes at the conifold
singularity \cite{Klebanov:1998hh}. For general $N$, its global symmetry is $\grSU(2)\times
\grSU(2)$ but for $N=2$ it becomes enhanced to $\grSU(4)$ \cite{Forcella:2008bb} (in this case the theory becomes 
equivalent to the BLG theory with an extra gauged $\grU(1)$ \cite{Aharony:2008ug}). For $N>2$ ABJM 
showed that this theory possesses $\superN=6$ supersymmetry \cite{Aharony:2008ug}.
In the $\superN=2$ superspace formulation, this means that,
for a specially chosen normalization of the superpotential,
the global symmetry is enhanced to $\grSU(4)_R$. We demonstrate explicitly how this symmetry enhancement
happens in terms of the component fields, once again due to a special cancellation involving F and D terms.

In \secref{sec:quivercircle} we consider a $\Integers_n$
orbifold of the ABJM theory that produces a $(\grU(N)\times \grU(N))^n$ Chern-Simons gauge theory. 
This theory possesses $\grSU(2)\times \grSU(2)$ R-symmetry, indicating that it has
$\superN=4$ supersymmetry. We propose that this theory describes $N$ M2-branes at a certain
$\Integers_n\times \Integers_{\tilde{k}}$ orbifold of $\Complex^4$, where $\tilde{k}$ is linearly related to the level $k$. Thus, this theory is conjectured to be dual, in the sense of \cite{Maldacena:1997re,Gubser:1998bc,Witten:1998qj}, to a certain $\Integers_n\times\Integers_{\tilde{k}}$ orbifold of $AdS_4\times S^7$. In \secref{sec:chiralquiver} we consider a different $\Integers_l$ orbifold of the ABJM theory that produces a family of chiral $(\grU(N)\times \grU(N))^l$ Chern-Simons gauge theories with $\superN=2$ supersymmetry and $\grSU(2)$ global symmetry. These theories are conjectured to be dual to $\Integers_l\times\Integers_{\tilde{k}}$ orbifolds of $AdS_4\times S^7$ that preserve the same symmetries.
In \secref{sec:supersu3} we deform the quartic superpotential
of the $\superN=8$ theory by an $\grSU(3)$ invariant mass term, creating RG flow 
to an $\superN=2$ superconformal gauge theory with a sextic superpotential. 
We conjecture that this new gauge theory is dual to the $\grU(1)_R\times \grSU(3)$ invariant extremum \cite{Warner:1983vz} of
the $\superN=8$ gauged supergravity. 
We also propose that the entire $\grSU(3)$ invariant RG flow in the Chern-Simons gauge theory
is dual to the M-theory description found in 
\cite{Ahn:2000aq,Corrado:2001nv,Johnson:2001ze,Ahn:2002eh}.

\section{Summary of BLG theory}
\label{sec:BL-review}

Here we review the BLG theory in van Raamsdonk's product gauge group formulation \cite{VanRaamsdonk:2008ft}, which rewrites it as a superconformal Chern-Simons theory with $\grSU(2)^2$ gauge group and bi-fundamental matter. It has a manifest global $\grSO(8)$ R-symmetry which shows that it is $\superN=8$ supersymmetric.

We use the following notation. Indices transforming under the first $\grSU(2)$ factor of the gauge group are $a,b,\ldots$, and for the second factor we use $\hat{a},\hat{b},\ldots$. Fundamental indices are written as superscript and anti-fundamental indices as subscript. Thus, the gauge and matter fields are $A^a{}_b$, $\hat{A}^{\hat{a}}{}_{\hat{b}}$, $X^a{}_{\hat{b}}$, and $\Psi^a{}_{\hat{b}}$. The conjugate fields have indices $(X^\dagger)^{\hat{a}}{}_b$ and $(\Psi^\dagger)^{\hat{a}}{}_b$. Most of the time, however, we will use matrix notation and suppress gauge indices. Lorentz indices are $\mu=0,1,2$ and the metric on the world volume is $g_{\mu\nu} = \diag(-1,+1,+1)$. $\grSO(8)$ vector indices are $I,J,\ldots$. The fermions a represented by 32-component Majorana spinors of $\grSO(1,10)$ subject to a chirality condition on the world-volume which leaves 16 real degrees of freedom. The $\grSO(1,10)$ spinor indices are generally omitted.

The action is then given by \cite{VanRaamsdonk:2008ft}
\be \label{eqn:BL-action}
  \Action \eq \int d^3x \tr \Bigsbrk{
    - (\deriD^\mu X^I)^\dagger \deriD_\mu X^I  +  i\bar{\Psi}^\dagger \Gamma^\mu \deriD_\mu \Psi \nl
    -\frac{2if}{3} \bar{\Psi}^\dagger \Gamma^{IJ} \bigbrk{X^I X^{J\dagger} \Psi + X^J \Psi^\dagger X^I + \Psi X^{I\dagger} X^J}
    -\frac{8f^2}{3} \tr X^{[I} X^{\dagger J} X^{K]} X^{\dagger[K} X^{J} X^{\dagger I]} \nl
    + \frac{1}{2f} \levi^{\mu\nu\lambda}(A_\mu \partial_\nu A_\lambda + \frac{2i}{3} A_\mu A_\nu A_\lambda)
    - \frac{1}{2f} \levi^{\mu\nu\lambda}(\hat{A}_\mu \partial_\nu \hat{A}_\lambda + \frac{2i}{3} \hat{A}_\mu \hat{A}_\nu \hat{A}_\lambda)
  }
\ee
where the covariant derivative is
\be \label{eqn:cov-derivative}
  \deriD_\mu X = \partial_\mu X + i A_\mu X - i X \hat{A}_\mu \; .
\ee
The Chern-Simons level $k$ is contained in
\be
  f = \frac{2\pi}{k} \; .
\ee

The bifundamental scalars $X^I$ are related to the original BLG variables $x_a^I$ with $\grSO(4)$ index $a$ through
\be \label{eqn:map-so4-su2su2}
  X^I = \Half (x_4^I \unit + i x_i^I \sigma^i) \; ,
\ee
where $\sigma^i$ are the Pauli matrices. It is important to note that the scalars satisfy the reality condition
\be \label{eqn:reality-condition}
  X^* = - \eps X \eps \; ,
\ee
where $\eps = i\sigma_2$. This condition can only be imposed for the gauge group $\grSU(2)^2$, which seems to present an obstacle for generalizing the theory to rank $N>2$. Recently, this obstacle was overcome by using complex bifundamental superfields \cite{Aharony:2008ug}. This will be reviewed in \secref{sec:N-N}. 

Finally we note the form of the $\grSU(2)^2$ gauge transformations
\begin{align} \label{eqn:gauge-trafo-components}
  A_\mu        & \rightarrow U A_\mu U^\dagger - i U \partial_\mu U^\dagger \; , &
  X            & \rightarrow U X \hat{U}^\dagger \; , \\ \nn
  \hat{A}_\mu  & \rightarrow \hat{U} \hat{A}_\mu \hat{U}^\dagger - i \hat{U} \partial_\mu \hat{U}^\dagger \; , &
  X^\dagger    & \rightarrow \hat{U} X^\dagger U^\dagger \; ,
\end{align}
where $U,\hat{U}\in\grSU(2)$.

\section{BLG theory in $\superN=2$ superspace}
\label{sec:BL-superspace}

In this section we will write the BLG theory \eqref{eqn:BL-action} in $\superN=2$ superspace. 
Of the $\grSO(8)_R$ symmetry this formalism leaves only the subgroup $\grU(1)_R\times \grSU(4)$ manifest. However, we will demonstrate how the $\grSO(8)$ R-symmetry is recovered when the action is expressed in terms of component fields. Our notations and many useful superspace identities are summarized in \appref{sec:notations}.

The gauge fields $A$ and $\hat{A}$ become components of two gauge vector superfields $\VV$ and $\hat{\VV}$. Their component expansions in Wess-Zumino gauge are
\be \label{eqn:V-components}
  \VV = 2i \, \theta \bar{\theta} \, \sigma(x)
    + 2 \, \theta\gamma^\mu\bar{\theta} \, A_\mu(x)
    + \sqrt{2} i \, \theta^2 \, \bar{\theta}\bar{\chi}(x) 
    - \sqrt{2} i \, \bar{\theta}^2 \, \theta \chi(x)
    + \theta^2 \, \bar{\theta}^2 \, \auxD(x)
\ee
and correspondingly for $\hat{\VV}$. Here $\sigma$ and $\auxD$ are auxiliary scalars, and $\chi$ and $\bar{\chi}$ are auxiliary fermions. The matter fields $X$ and $\Psi$ are accommodated in chiral superfields $\ZZ$ and anti-chiral superfields $\bar{\ZZ}$ which transform in the fundamental and anti-fundamental representation of $\grSU(4)$, respectively. Their $\grSU(4)$ indices $\ZZ^A$ and $\bar{\ZZ}_A$ will often be suppressed. The component expansions are
\be \label{eqn:Z-components}
  \ZZ \eq Z(x_L)
       + \sqrt{2} \theta\zeta(x_L)
       + \theta^2 \, F(x_L) \; , \\ \label{eqn:barZ-components}
  \bar{\ZZ} \eq Z^\dagger(x_R)
       - \sqrt{2} \bar{\theta}\zeta^\dagger(x_R)
       - \bar{\theta}^2 \, F^\dagger(x_R) \; .
\ee
The scalars $Z$ are complex combinations of the BLG scalars
\be \label{eqn:Z=XIX}
  Z^A = X^A + i X^{A+4} \qquad\mbox{for $A=1,\ldots,4$} \; .
\ee
We define two operations which conjugate the $\grSU(2)$ representations and the $\grSU(4)$ representation, respectively, as\footnote{We should caution that the bar denoting the anti-chiral superfield $\bar{\ZZ}$ is just a label and does \emph{not} mean that the component fields are conjugated by \eqref{eqn:def-bar}. In fact, the components of $\bar{\ZZ}$ are the hermitian conjugates, see \eqref{eqn:barZ-components}.}
\begin{align}
  Z^{\ddagger A} &:= - \eps (Z^A)^\trans \eps = X^{\dagger A} + i X^{\dagger A+4} \; , \label{eqn:def-ddagger} \\
  \bar{Z}_A      &:= - \eps (Z^A)^*      \eps = X^A           - i X^{A+4} \; . \label{eqn:def-bar}
\end{align}
Separating these two operations in possible only for gauge group $\grSU(2)^2$, since for gauge groups of higher rank there is no reality condition analogous to \eqref{eqn:reality-condition}. In these cases only the combined action, which is the hermitian conjugate $Z^\dagger = \bar{Z}^\ddagger$, makes sense. The possibility to conjugate the $\grSU(4)$ representation independently from the $\grSU(2)^2$ representation allows us to invert \eqref{eqn:Z=XIX}:
\be
  X^A     = \tfrac{1}{2} \bigbrk{ Z^A + \bar{Z}_A }
  \comma
  X^{A+4} = \tfrac{1}{2i} \bigbrk{ Z^A - \bar{Z}_A }
  \; .
\ee

The superspace action $\Action = \Action_{\mathrm{CS}} + \Action_{\mathrm{mat}} + \Action_{\mathrm{pot}}$ consists of a Chern-Simons part, a matter part and a superpotential given by
\be
  \Action_{\mathrm{CS}} \eq -i K \int d^3x\,d^4\theta \int_0^1 dt\: \tr \Bigsbrk{ \VV \bar{D}^\alpha \Bigbrk{ e^{t \VV} D_\alpha e^{-t \VV} } - \hat{\VV} \bar{D}^\alpha \Bigbrk{ e^{t \hat{\VV}} D_\alpha e^{-t \hat{\VV}} } } \; , \label{eqn:CS-action} \\
  \Action_{\mathrm{mat}} \eq - \int d^3x\,d^4\theta\: \tr \bar{\ZZ}_A e^{-\VV} \ZZ^A e^{\hat{\VV}} \; , \label{eqn:mat-action-2-2} \\
  \Action_{\mathrm{pot}} \eq L \int d^3x\,d^2\theta\: \superW(\ZZ) + L \int d^3x\,d^2\bar{\theta}\: \bar{\superW}(\bar{\ZZ}) \label{eqn:superpot-2-2}
\ee
with
\be \label{eqn:superpot}
  \superW       = \frac{1}{4!} \levi_{ABCD} \tr \ZZ^A \ZZ^{\ddagger B} \ZZ^C \ZZ^{\ddagger D}
  \comma
  \bar{\superW} = \frac{1}{4!} \levi^{ABCD} \tr \bar{\ZZ}_A \bar{\ZZ}_B^\ddagger \bar{\ZZ}_C \bar{\ZZ}_D^\ddagger
  \; .
\ee
In terms of $\grSO(4)$ variables, $\mathcal{Z}_a$, which are related to the $\grSU(2)^2$ fields according to 
\eqref{eqn:map-so4-su2su2}, it assumes the form 
\be \label{eqn:superpot4}
  \superW = -\frac{1}{8\cdot 4!} \; \levi_{ABCD} \levi^{abcd} \ZZ^A_a \ZZ^B_b \ZZ^C_c \ZZ^D_d \; .
\ee
This superpotential possesses only a $\grU(1)_R\times \grSU(4)$ 
global symmetry as opposed to the $\grSO(8)_R$ symmetry of the BLG theory. We will show in the following that when the normalization constants $K$ and $L$ are related as $K = \frac{1}{L}$, then the R-symmetry of the model is enhanced to $\grSO(8)$. If we furthermore set $L=4f$, we recover precisely the action \eqref{eqn:BL-action}.

The gauge transformations are given by \cite{Ivanov:1991fn}
\be
  e^{t \VV}       \rightarrow e^{i\Lambda} e^{t \VV} e^{-i\bar{\Lambda}} \; , \;\;
  e^{t \hat{\VV}} \rightarrow e^{i\hat{\Lambda}} e^{t \hat{\VV}} e^{-i\hat{\bar{\Lambda}}} \; , \;\;
  \ZZ             \rightarrow e^{i\Lambda} \ZZ e^{-i\hat{\Lambda}} \; , \;\;
  \bar{\ZZ}       \rightarrow e^{i\hat{\bar{\Lambda}}} \bar{\ZZ} e^{-i\bar{\Lambda}} \; ,
\ee
where the parameters $\Lambda,\hat{\Lambda}$ and $\bar{\Lambda},\hat{\bar{\Lambda}}$ are chiral and anti-chiral superfields, respectively. Their $t$ dependence is determined by consistency of the transformation law for $\VV$ and $\hat{\VV}$. In order to preserve the WZ gauge, these fields have to be simply
\be
  \Lambda = \lambda(x_L) \comma
  \bar{\Lambda} = \lambda(x_R) \comma
  \hat{\Lambda} = \hat{\lambda}(x_L) \comma
  \hat{\bar{\Lambda}} = \hat{\lambda}(x_R) \; .
\ee
with $\lambda$ and $\hat{\lambda}$ real. These transformations reduce to the ones given in \eqref{eqn:gauge-trafo-components} when we set $U(x)\equiv e^{i\lambda(x)}$ and $\hat{U}(x)\equiv e^{i\hat{\lambda}(x)}$.

\paragraph{Expressions in components.} We will now show that the above superspace action describes the BLG theory by expanding it into component fields. The Chern-Simons action then reads
\be \label{eqn:CS-action-components}
  \Action_{\mathrm{CS}} \eq
  K \int d^3x\: \tr \Bigsbrk{
   2 \levi^{\mu\nu\lambda}(A_\mu \partial_\nu A_\lambda + \frac{2i}{3} A_\mu A_\nu A_\lambda)
  -2 \levi^{\mu\nu\lambda}(\hat{A}_\mu \partial_\nu \hat{A}_\lambda + \frac{2i}{3} \hat{A}_\mu \hat{A}_\nu \hat{A}_\lambda)
\nl \qquad\qquad\qquad
  + 2i \bar{\chi} \chi - 2i \hat{\bar{\chi}} \hat{\chi}
  - 4 \auxD \sigma + 4 \hat{\auxD} \hat{\sigma}
  }
\ee
and the matter action becomes
\be \label{eqn:mat-action-components}
  \Action_{\mathrm{mat}} \eq \int d^3x\: \tr \Bigsbrk{
   - (\deriD_\mu Z)^\dagger \deriD^\mu Z
   - i \zeta^\dagger \slashed{\deriD} \zeta
   + F^\dagger F
   + Z^\dagger \auxD Z
   - Z^\dagger Z \hat{\auxD}
\nl \qquad\qquad\quad
   + i Z^\dagger \chi \zeta
   + i \zeta^\dagger \bar{\chi} Z
   - i Z^\dagger \zeta \hat{\chi}
   - i \zeta^\dagger Z \hat{\bar{\chi}}
\nl[2mm] \qquad\qquad\quad
   - Z^\dagger \sigma^2 Z
   - Z^\dagger Z \hat{\sigma}^2
   + 2Z^\dagger \sigma Z \hat{\sigma}
   - i\zeta^\dagger \sigma \zeta
   + i\zeta^\dagger \zeta \hat{\sigma}
 } \; .
\ee
The gauge covariant derivative is defined in \eqref{eqn:cov-derivative}. Let us remind that our notation suppresses indices in ``standard positions''\footnote{The standard position of an index is defined when the field is introduced and those for spinor indices are explained in \appref{sec:notations}.}, e.g.
\be
 \tr Z^\dagger \chi \zeta 
 \;\equiv\;
 \tr Z^\dagger_A \chi^\alpha \zeta_\alpha^A
 \;\equiv\;
 (Z^\dagger_A)^{\hat{a}}{}_b (\chi^\alpha)^b{}_c (\zeta_\alpha^A)^c{}_{\hat{a}}
 \; .
\ee
The superpotential contains the following interactions of the component fields
\be
  \Action_{\mathrm{pot}} \eq - \frac{L}{12} \int d^3x\: \tr \Bigsbrk{
\levi_{ABCD} \bigbrk{
     \zeta^A \zeta^{\ddagger B} Z^C Z^{\ddagger D}
   - \zeta^{\ddagger A} \zeta^B Z^{\ddagger C} Z^D
   + \zeta^A Z^{\ddagger B} \zeta^C Z^{\ddagger D}
   }
\nln && \hspace{24mm}
+ \levi^{ABCD} \bigbrk{
     \bar{\zeta}^\ddagger_A \bar{\zeta}_B \bar{Z}^\ddagger_C \bar{Z}_D
   - \bar{\zeta}_A \bar{\zeta}^\ddagger_B \bar{Z}_C \bar{Z}^\ddagger_D
   + \bar{\zeta}^\ddagger_A \bar{Z}_B \bar{\zeta}^\ddagger_C \bar{Z}_D
  }
\nln[1mm] && \hspace{24mm}
   + 2 \levi_{ABCD} F^A Z^{\ddagger B} Z^C Z^{\ddagger D}
   - 2 \levi^{ABCD} \bar{F}^\ddagger_A \bar{Z}_B \bar{Z}^\ddagger_C \bar{Z}_D
   } \; .
\ee

\paragraph{Integrating out auxiliary fields.} The fields $\auxD$ and $\hat{\auxD}$ are Lagrange multipliers for the constraints
\be
  \sigma^n       = \frac{1}{4K} \tr t^n Z Z^\dagger
  \comma
  \hat{\sigma}^n = \frac{1}{4K} \tr t^n Z^\dagger Z
  \; ,
\ee
where $t^n$ are the generators of $\grSU(2)$ normalized as described in \appref{sec:notations}. The equations of motion for the $\chi$'s are
\begin{align}
  \chi^n             & = -\frac{1}{2K} \tr t^n Z \zeta^\dagger \; , &
  \bar{\chi}^n       & = -\frac{1}{2K} \tr t^n \zeta Z^\dagger \; , \\
  \hat{\chi}^n       & = -\frac{1}{2K} \tr t^n \zeta^\dagger Z \; , &
  \hat{\bar{\chi}}^n & = -\frac{1}{2K} \tr t^n Z^\dagger \zeta \; ,
\end{align}
and the ones for $F$ are
\be
  F^A         =  -\frac{L}{6} \levi^{ABCD} \bar{Z}_B \bar{Z}^\ddagger_C \bar{Z}_D 
  \comma
  F^\dagger_A =  +\frac{L}{6} \levi_{ABCD} Z^{\ddagger B} Z^C Z^{\ddagger D}
  \; .
\ee
Using these relations one finds the following action
\be
  \Action \eq \int d^3x\: \Bigsbrk{ 
    2 K \levi^{\mu\nu\lambda} \tr \bigbrk{
           A_\mu \partial_\nu A_\lambda + \tfrac{2i}{3} A_\mu A_\nu A_\lambda
         - \hat{A}_\mu \partial_\nu \hat{A}_\lambda - \tfrac{2i}{3} \hat{A}_\mu \hat{A}_\nu \hat{A}_\lambda
         }
\nl \qquad\qquad\quad
    - \tr (\deriD_\mu Z)^\dagger \deriD^\mu Z
    - i \tr \zeta^\dagger \slashed{\deriD} \zeta 
    - V_{\mathrm{ferm}} - V_{\mathrm{bos}}
} \; .
\ee
The quartic terms $V_{\mathrm{ferm}}$ are interactions between fermions and bosons, and the sextic terms $V_{\mathrm{bos}}$ are interactions between bosons only. Separated according to their origin we have
\be
  V_D^{\mathrm{ferm}} \eq 
     \frac{i}{4K} \tr \Bigsbrk{
     \zeta^A \zeta_A^\dagger Z^B Z_B^\dagger
   - \zeta_A^\dagger \zeta^A Z_B^\dagger Z^B  
   + 2 \zeta^A Z_A^\dagger Z^B \zeta_B^\dagger
   - 2 Z_A^\dagger \zeta^A \zeta_B^\dagger Z^B
   } \; , \\[2mm]
  V_F^{\mathrm{ferm}} \eq
  \frac{L}{12} \levi_{ABCD} \tr \Bigsbrk{
     \zeta^A \zeta^{\ddagger B} Z^C Z^{\ddagger D}
   - \zeta^{\ddagger A} \zeta^B Z^{\ddagger C} Z^D
   + \zeta^A Z^{\ddagger B} \zeta^C Z^{\ddagger D}
   }
\nl
+ \frac{L}{12} \levi^{ABCD} \tr \Bigsbrk{
     \bar{\zeta}^\ddagger_A \bar{\zeta}_B \bar{Z}^\ddagger_C \bar{Z}_D
   - \bar{\zeta}_A \bar{\zeta}^\ddagger_B \bar{Z}_C \bar{Z}^\ddagger_D
   + \bar{\zeta}^\ddagger_A \bar{Z}_B \bar{\zeta}^\ddagger_C \bar{Z}_D
  }
\ee
and
\be
  V_D^{\mathrm{bos}} \eq 
     \frac{1}{16K^2} \tr \Bigsbrk{
     Z^A Z_A^\dagger Z^B Z_B^\dagger Z^C Z_C^\dagger
   + Z_A^\dagger Z^A Z_B^\dagger Z^B Z_C^\dagger Z^C
   - 2 Z_A^\dagger Z^B Z_B^\dagger Z^A Z_C^\dagger Z^C
   } \; , \\[2mm]
  V_F^{\mathrm{bos}} \eq
     - \frac{L^2}{36} \levi_{ABCG} \levi^{DEFG} \tr
     Z^{\ddagger A} Z^B Z^{\ddagger C} \bar{Z}_D \bar{Z}^\ddagger_E \bar{Z}_F
\; .
\ee
When substituting in \eqref{eqn:Z=XIX} we find that for $K = \frac{1}{L}$ all sextic interactions can be joined together to
\be
  V^{\mathrm{bos}} = \frac{L^2}{6} \tr X^{[I} X^{\dagger J} X^{K]} X^{\dagger[K} X^{J} X^{\dagger I]} \; .
\ee
Furthermore setting $L=4f$, this is precisely the scalar potential of the BLG theory \eqref{eqn:BL-action}. With this choice also the other coefficients match exactly.

\section{ABJM $\grU(N)^2$ gauge theory in superspace}
\label{sec:N-N}

As remarked in \secref{sec:BL-review}, it is not obvious how to generalize van Raamsdonk's formulation of the BLG theory to higher rank gauge groups. This difficulty is also evident in our superspace formulation, since the manifestly $\grSU(4)$ invariant superpotential is gauge invariant only for $\grSU(2)\times \grSU(2)$ gauge theory. A way forward is the recently proposed generalization due to ABJM \cite{Aharony:2008ug}.

Their key idea for the generalization is to give up the manifest global $\grSU(4)$ invariance by forming 
the following complex combinations of the bifundamental fields:
\begin{align} \label{eqn:su4-so8}
  Z^1 & = X^1          + i X^5 \; , &
  W_1 & = X^{3\dagger} + i X^{7\dagger} \; , \\
  Z^2 & = X^2          + i X^6 \; , &  W_2 & = X^{4\dagger} + i X^{8\dagger} \; .
\end{align}
Promoting these fields to chiral superfields, the superpotential of the BLG
theory \eqref{eqn:superpot} may be written as
\cite{Aharony:2008ug} 
\be \label{eqn:superpot-N-N}
  \Action_{\mathrm{pot}} =
     L \int d^3x\,d^2\theta\: \superW(\ZZ,\WW)
   + L \int d^3x\,d^2\bar{\theta}\: \bar{\superW}(\bar{\ZZ},\bar{\WW})
\ee 
with 
\be
  \superW    = \frac{1}{4} \levi_{AC}\levi^{BD} \tr \ZZ^A \WW_B \ZZ^C \WW_D
  \comma
  \bar{\superW} = \frac{1}{4} \levi^{AC}\levi_{BD} \tr \bar{\ZZ}_A \bar{\WW}^B \bar{\ZZ}_C \bar{\WW}^D
  \; .
\ee
This form of the superpotential is exactly the same as for the theory on D3-branes on the conifold \cite{Klebanov:1998hh} and it generalizes readily to $\grSU(N)\times \grSU(N)$ gauge group. This superpotential has a global symmetry $\grSU(2)\times \grSU(2)$ and also a ``baryonic'' $\grU(1)$ symmetry
\be \label{eqn:baryonic}
  \ZZ^A \rightarrow e^{i\alpha} \ZZ^A
  \comma
  \WW_B \rightarrow e^{-i\alpha} \WW_B
\; .
\ee 
In the $3+1$ dimensional case this symmetry is originally gauged, but far in the IR it 
becomes global \cite{Klebanov:1998hh}. However, in the present $2+1$ dimensional example this 
does not happen, so it is natural to add it to the gauge symmetry \cite{Aharony:2008ug}. 
Including also the trivial neutral $\grU(1)$, we thus find the $\grU(N)\times \grU(N)$ Chern-Simons gauge 
theory at level $k$. The gauging of the symmetry \eqref{eqn:baryonic} seems important for obtaining the 
correct M-theory interpretation for arbitrary $k$ and $N$ \cite{Aharony:2008ug}. 
Since this symmetry corresponds to simultaneous rotation of the 4 complex coordinates of $\Complex^4$ 
transverse to the M2-branes, this space actually turns into an orbifold $\Complex^4/\Integers_k$ 
\cite{Aharony:2008ug}. Because of this gauging, even for $N=2$ the ABJM theory is slightly different from
the BLG theory.

\bigskip

Let us summarize the properties of the ABJM theory \cite{Aharony:2008ug} and explicitly prove that its $\grU(1)_R\times \grSU(2)\times\grSU(2)$ global symmetry becomes enhanced to $\grSU(4)_R$. The fields $\ZZ$ and $\WW$ transform in the $(\rep{2},\rep{1})$ and the $(\rep{1},\rep{\bar{2}})$ of the global $\grSU(2)^2$ and in the $(\rep{N},\rep{\bar{N}})$ and the $(\rep{\bar{N}},\rep{N})$ of the gauge group $\grU(N)^2$, respectively. We use the following conventions for $\grSU(2)^2$ indices: $\ZZ^A$, $\bar{\ZZ}_A$, $\WW_A$, $\bar{\WW}^A$ and for $\grU(N)^2$ indices: $\ZZ^a{}_{\hat{a}}$, $\bar{\ZZ}^{\hat{a}}{}_a$, $\WW^{\hat{a}}{}_a$, $\bar{\WW}^a{}_{\hat{a}}$. The gauge superfields have indices $\VV^a{}_b$ and $\hat{\VV}^{\hat{a}}{}_{\hat{b}}$. The component fields for $\ZZ$, $\bar{\ZZ}$ and $\VV$ are as previously in \eqref{eqn:Z-components}, \eqref{eqn:barZ-components} and \eqref{eqn:V-components}. The components of $\WW$ and $\bar{\WW}$ will be denoted by
\be \label{eqn:W-components}
  \WW \eq W(x_L)
       + \sqrt{2} \theta\omega(x_L)
       + \theta^2 \, G(x_L) \; , \\ \label{eqn:barW-components}
  \bar{\WW} \eq W^\dagger(x_R)
       - \sqrt{2} \bar{\theta}\omega^\dagger(x_R)
       - \bar{\theta}^2 \, G^\dagger(x_R) \; .
\ee
The Chern-Simons action is formally unaltered \eqref{eqn:CS-action}, the matter part \eqref{eqn:mat-action-2-2} splits into 
\be
  \Action_{\mathrm{mat}} = \int d^3x\,d^4\theta\: \tr \Bigsbrk{ - \bar{\ZZ}_A e^{-\VV} \ZZ^A e^{\hat{\VV}} - \bar{\WW}^A e^{-\hat{\VV}} \WW_A e^{\VV} } \; ,
\ee
and the superpotential is given by \eqref{eqn:superpot-N-N}. The symmetry enhancement to $\grSU(4)_R$ requires the normalization constants in \eqref{eqn:CS-action} and \eqref{eqn:superpot-N-N} to be related as $K=\frac{1}{L}$.

\paragraph{Expressions in components.} The component form of the Chern-Simons action has been computed in \eqref{eqn:CS-action-components} and the matter action involving $\ZZ$ looks identical to \eqref{eqn:mat-action-components} where now $Z,\zeta,F$ have only two components. The matter action for $\WW$ is analogously given by
\be 
  \Action_{\mathrm{mat}}^\WW \eq \int d^3x\: \tr \Bigsbrk{
   - (\deriD_\mu W)^\dagger \deriD^\mu W
   - i \omega^\dagger \slashed{\deriD} \omega
   + G^\dagger G
   + W^\dagger \hat{\auxD} W
   - W^\dagger W \auxD
\nl \qquad\qquad\quad
   + i W^\dagger \hat{\chi} \omega
   + i \omega^\dagger \hat{\bar{\chi}} W
   - i W^\dagger \omega \chi
   - i \omega^\dagger W \bar{\chi}
\nl[2mm] \qquad\qquad\quad
   - W^\dagger \hat{\sigma}^2 W
   - W^\dagger W \sigma^2
   + 2W^\dagger \hat{\sigma} W \sigma
   - i\omega^\dagger \hat{\sigma} \omega
   + i\omega^\dagger \omega \sigma } \; ,
\ee
where $\deriD_\mu W = \partial_\mu W + i \hat{A}_\mu W - i W A_\mu$. The superpotential expands to
\be
  \Action_{\mathrm{pot}} \eq \frac{L}{4} \int d^3x\: \tr \Bigsbrk{ \levi_{AC} \levi^{BD} \Bigbrk{
     2 F^A W_B Z^C W_D
   + 2 Z^A W_B Z^C G_D
\\ && \qquad\qquad
   - 2 \zeta^A W_B Z^C \omega_D
   - 2 \zeta^A \omega_B Z^C W_D
   -   Z^A \omega_B Z^C \omega_D
   -   \zeta^A W_B \zeta^C W_D
   }
\nln && \qquad 
- \levi^{AC} \levi_{BD} \Bigbrk{
     2 F^\dagger_A W^{\dagger B} Z^\dagger_C W^{\dagger D}
   + 2 Z^\dagger_A W^{\dagger B} Z^\dagger_C G^{\dagger D}
\nln && \qquad\qquad
   + 2 \zeta^\dagger_A W^{\dagger B} Z^\dagger_C \omega^{\dagger D}
   + 2 \zeta^\dagger_A \omega^{\dagger B} Z^\dagger_C W^{\dagger D}
   +   Z^\dagger_A \omega^{\dagger B} Z^\dagger_C \omega^{\dagger D}
   +   \zeta^\dagger_A W^{\dagger B} \zeta^\dagger_C W^{\dagger D}
  } } \; .
  \nn
\ee

\paragraph{Integrating out auxiliary fields.} The auxiliary fields can be replaced by means of the following equations:
\begin{align}
  \sigma^n       & = \frac{1}{4K} \tr T^n \bigbrk{ Z Z^\dagger - W^\dagger W } \; , &
  \hat{\sigma}^n & = \frac{1}{4K} \tr T^n \bigbrk{ Z^\dagger Z - W W^\dagger } \; , \\[3mm]
  \chi^n             & = -\frac{1}{2K} \tr T^n \bigbrk{ Z \zeta^\dagger - \omega^\dagger W } \; , &
  \bar{\chi}^n       & = -\frac{1}{2K} \tr T^n \bigbrk{ \zeta Z^\dagger - W^\dagger \omega } \; , \\
  \hat{\chi}^n       & = -\frac{1}{2K} \tr T^n \bigbrk{ \zeta^\dagger Z - W \omega^\dagger } \; , &
  \hat{\bar{\chi}}^n & = -\frac{1}{2K} \tr T^n \bigbrk{ Z^\dagger \zeta - \omega W^\dagger } \; , \\[3mm]
  F^A           & = +\frac{L}{2} \levi^{AC} \levi_{BD} W^{\dagger B} Z^\dagger_C W^{\dagger D} \; , &
  G_A           & = -\frac{L}{2} \levi_{AC} \levi^{BD} Z^\dagger_B W^{\dagger C} Z^\dagger_D \; , \label{eqn:auxF} \\
  F^\dagger_A   & = -\frac{L}{2} \levi_{AC} \levi^{BD} W_B Z^C W_D \; , &
  G^{\dagger A} & = +\frac{L}{2} \levi^{AC} \levi_{BD} Z^B W_C Z^D \; . \label{eqn:auxG}
\end{align}
Then the total action reads
\be
  \Action \eq \int d^3x\: \Bigsbrk{ 
    2 K \levi^{\mu\nu\lambda} \tr \bigbrk{
        A_\mu \partial_\nu A_\lambda + \tfrac{2i}{3} A_\mu A_\nu A_\lambda
        - \hat{A}_\mu \partial_\nu \hat{A}_\lambda - \tfrac{2i}{3} \hat{A}_\mu \hat{A}_\nu \hat{A}_\lambda
      }
\nl \hspace{20mm}
    - \tr (\deriD_\mu Z)^\dagger \deriD^\mu Z
    - \tr (\deriD_\mu W)^\dagger \deriD^\mu W
    - i \tr \zeta^\dagger \slashed{\deriD} \zeta 
    - i \tr \omega^\dagger \slashed{\deriD} \omega
\nl[1mm] \hspace{20mm}
    - V_{\mathrm{ferm}} - V_{\mathrm{bos}}
}
\ee
with the potentials
\be
  V_D^{\mathrm{ferm}} \eq 
     \frac{i}{4K} \tr \Bigsbrk{
     \bigbrk{\zeta^A \zeta_A^\dagger - \omega^{\dagger A} \omega_A} \bigbrk{Z^B Z_B^\dagger - W^{\dagger B} W_B}
   - \bigbrk{\zeta_A^\dagger \zeta^A - \omega_A \omega^{\dagger A}} \bigbrk{Z_B^\dagger Z^B - W_B W^{\dagger B}} 
   }
\nl
   + \frac{i}{2K} \tr \Bigsbrk{
     \bigbrk{\zeta^A Z_A^\dagger - W^{\dagger A} \omega_A} \bigbrk{Z^B \zeta_B^\dagger - \omega^{\dagger B} W_B}
   - \bigbrk{Z_A^\dagger \zeta^A - \omega_A W^{\dagger A}} \bigbrk{\zeta_B^\dagger Z^B - W_B \omega^{\dagger B}}
   } \; , \nn \\[2mm]
  V_F^{\mathrm{ferm}} \eq 
     \frac{L}{4} \levi_{AC} \levi^{BD} \tr \Bigsbrk{
     2 \zeta^A W_B Z^C \omega_D
   + 2 \zeta^A \omega_B Z^C W_D
   +   Z^A \omega_B Z^C \omega_D
   +   \zeta^A W_B \zeta^C W_D
   }
\nl
   + \frac{L}{4} \levi^{AC} \levi_{BD} \tr \Bigsbrk{
     2 \zeta^\dagger_A W^{\dagger B} Z^\dagger_C \omega^{\dagger D} 
   + 2 \zeta^\dagger_A \omega^{\dagger B} Z^\dagger_C W^{\dagger D}
   +   Z^\dagger_A \omega^{\dagger B} Z^\dagger_C \omega^{\dagger D}
   +   \zeta^\dagger_A W^{\dagger B} \zeta^\dagger_C W^{\dagger D}
   } \nn
\ee
and
\be
  V_D^{\mathrm{bos}} \eq 
     \frac{1}{16K^2} \tr \Bigsbrk{
     \bigbrk{Z^A Z_A^\dagger + W^{\dagger A} W_A} \bigbrk{Z^B Z_B^\dagger - W^{\dagger B} W_B} \bigbrk{Z^C Z_C^\dagger - W^{\dagger C} W_C}
\nl \hspace{17mm}
   + \bigbrk{Z_A^\dagger Z^A + W_A W^{\dagger A}} \bigbrk{Z_B^\dagger Z^B - W_B W^{\dagger B}} \bigbrk{Z_C^\dagger Z^C - W_C W^{\dagger C}}
\nl[1mm] \hspace{17mm}
   - 2 Z_A^\dagger \bigbrk{Z^B Z_B^\dagger - W^{\dagger B} W_B} Z^A \bigbrk{Z_C^\dagger Z^C - W_C W^{\dagger C}}
\nl \hspace{17mm}
   - 2 W^{\dagger A} \bigbrk{Z_B^\dagger Z^B - W_B W^{\dagger B}} W_A \bigbrk{Z^C Z_C^\dagger - W^{\dagger C} W_C}
   } \; , \nn \\[2mm]
  V_F^{\mathrm{bos}} \eq
   - \frac{L^2}{4} \tr \Bigsbrk{
     W^{\dagger A} Z_B^\dagger W^{\dagger C} W_A Z^B W_C
   - W^{\dagger A} Z_B^\dagger W^{\dagger C} W_C Z^B W_A
\nl \hspace{13mm}
   + Z_A^\dagger W^{\dagger B} Z_C^\dagger Z^A W_B Z^C
   - Z_A^\dagger W^{\dagger B} Z_C^\dagger Z^C W_B Z^A
   } \; . \nn
\ee

Let us note that $V_F^{\mathrm{bos}}$ and $V_D^{\mathrm{bos}}$ are separately non-negative.\footnote{We thank John Schwarz for useful discussions on this issue.} Indeed, the F-term contribution is related to the superpotential $\superW$ through
\be 
  V_F^{\mathrm{bos}} = \lrabs{\frac{\partial \superW}{\partial Z^A}}^2
                     + \lrabs{\frac{\partial \superW}{\partial W_A}}^2 
                     = \tr \bigsbrk{ F_A^\dagger  F^A + G^{\dagger A} G_A } \; ,
\ee
with $F^A$ and $G_A$ from (\ref{eqn:auxF},\ref{eqn:auxG}). The D-term contribution may be written as
\be
  V_D^{\mathrm{bos}} = \tr \bigsbrk{ N_A^\dagger  N^A + M^{\dagger A} M_A } \; ,
\ee
where $N^A = \sigma Z^A - Z^A \hat \sigma$ and $M_A = \hat\sigma W_A - W_A \sigma$. Thus, the total bosonic potential vanishes if and only if
\be
  F^A = G_A = N^A = M_A = 0 \; .
\ee

\paragraph{$\grSU(4)$ invariance.} If the coefficients of the Chern-Simons action and the superpotential are related by $K = \frac{1}{L}$, then the R-symmetry of the theory is enhanced to $\grSU(4)$.\footnote{This $\grSU(4)_R$ symmetry should not be confused with the global $\grSU(4)$ of the BLG theory. The latter is not manifest in the ABJM theory, but should nevertheless be present for $k=1$ and $k=2$ \cite{Aharony:2008ug}.} In order to make this symmetry manifest we combine the $\grSU(2)$ fields $Z$ and $W$ into a fundamental and anti-fundamental representation of $\grSU(4)$ as
\be
  Y^A = \{ Z^A, W^{\dagger A} \}
  \comma
  Y_A^\dagger = \{ Z_A^\dagger, W_A \} \; ,
\ee
where the index $A$ on the left hand side now runs from 1 to 4. Then the potential can be written as \cite{Aharony:2008ug}
\be \label{eqn:NN-Vbos}
   V^{\mathrm{bos}} \eq - \frac{L^2}{48} \tr \Bigsbrk{
          Y^A Y_A^\dagger Y^B Y_B^\dagger Y^C Y_C^\dagger 
      +   Y_A^\dagger Y^A Y_B^\dagger Y^B Y_C^\dagger Y^C
\nl\hspace{11mm}
      + 4 Y^A Y_B^\dagger Y^C Y_A^\dagger Y^B Y_C^\dagger 
      - 6 Y^A Y_B^\dagger Y^B Y_A^\dagger Y^C Y_C^\dagger 
   } \; .
\ee
The fermions have to be combined as follows
\be
  \psi_A = \{ \levi_{AB} \zeta^B \, e^{-i\pi/4}, -\levi_{AB} \omega^{\dagger B} \, e^{i\pi/4} \}
  \comma
  \psi^{A\dagger} = \{ -\levi^{AB} \zeta_B^\dagger \, e^{i\pi/4}, \levi^{AB} \omega_B \, e^{-i\pi/4} \} \; ,
\ee
and we can write fermionic interactions in the manifestly $\grSU(4)$ invariant way:
\be
   V^{\mathrm{ferm}} \eq \frac{i L}{4} \tr \Bigsbrk{
        Y_A^\dagger Y^A \psi^{B\dagger} \psi_B
      - Y^A Y_A^\dagger \psi_B \psi^{B\dagger}
      + 2 Y^A Y_B^\dagger \psi_A \psi^{B\dagger}
      - 2 Y_A^\dagger Y^B \psi^{A\dagger} \psi_B
\nl\hspace{10mm}
      - \levi^{ABCD} Y_A^\dagger \psi_B Y_C^\dagger \psi_D
      + \levi_{ABCD} Y^A \psi^{B\dagger} Y^C \psi^{D\dagger}
      } \; .
\ee
Thus, the $\grU(1)_R\times\grSU(2)\times\grSU(2)$ global symmetry is enhanced to $\grSU(4)_R$ symmetry, with the $\grU(1)_R$ corresponding to the generator $\half\diag(1,1,-1,-1)$. This shows that the theory in general possesses $\superN=6$ supersymmetry. 

In \cite{Aharony:2008ug} it was proposed that this $\grU(N)\times \grU(N)$ Chern-Simons theory at 
level $k$ describes the world volume of $N$ coincident M2-branes placed at the 
$\Integers_k$ orbifold of $\Complex^4$ where the action on the 4 complex 
coordinates\footnote{Let us note that these coordinates are not the same as the 
complex coordinates $z^A$ natural for the superspace formulation of 
BLG theory in \secref{sec:BL-superspace}. 
They are related through $y^1=z^1, y^2=z^2, y^3={\bar z}^3, y^4={\bar z}^4$.} 
is $y^A\rightarrow e^{2\pi i/k} y^A$. This action preserves the $\grSU(4)$ symmetry that rotates them, which in the gauge theory is realized as the R-symmetry. The $\superN=6$ supersymmetry of this orbifold can be checked as follows. The generator of $\Integers_k$ acts on the spinors of $\grSO(8)$ as
\be
\Psi \rightarrow e^{2\pi i (s_1 + s_2 + s_3 + s_4)/k} \Psi
\; ,
\ee
where $s_i=\pm 1/2$ are the spinor weights. The chirality projection implies that the sum of all $s_i$ must be even, producing an 8-dimensional representation. The spinors that are left invariant by the orbifold have $\sum_{i=1}^4 s_i = 0 \;(\mathrm{mod}\, k)$. This selects 6 out of the 8 spinors; therefore, the theory on M2-branes has 12 supercharges in perfect agreement with the Chern-Simons gauge theory with general level $k$.\footnote{For $k=1$ and $k=2$ there is further enhancement to $\superN=8$ supersymmetry, which is subtle in the gauge theory \cite{Aharony:2008ug}.} This is one of the reasons why the theory reviewed in this section was conjectured to be dual to M-theory on $AdS_4\times S^7/\Integers_k$ with $N$ units of flux \cite{Aharony:2008ug}.

\section{Non-chiral orbifold gauge theories }

\label{sec:quivercircle}

The results reviewed in the previous sections clearly represent major progress in understanding 
AdS$_4$/CFT$_3$ duality. In this section we make the first steps towards generalizing them. 
We will consider a further $\Integers_n$ projection of the basic $\grSU(4)$-invariant $\Integers_k$ orbifold 
reviewed in \secref{sec:N-N}.\footnote{A $\Integers_2$ orbifold of the BLG theory was also studied in \cite{Fuji:2008yj}. 
This orbifold is contained in our construction as a special case.} The $\Integers_n$ action is 
\be
y^1\rightarrow e^{2\pi i/n} y^1\; , \quad
y^2\rightarrow  y^2\; , \quad
y^3\rightarrow  e^{2\pi i/n} y^3\; , \quad
y^4\rightarrow  y^4\; .
\ee
This rotates the spinors of $\grSO(8)$ by the phase $e^{2\pi i(s_1 + s_3)/n}$. Thus, the spinors left
invariant by the combined $\Integers_k$ and $\Integers_n$ actions have $s_1 + s_3 = s_2 + s_4 = 0$. 
There are 4 such spinors corresponding to $\superN=4$ supersymmetry. The orbifold action preserves $\grSU(2)\times\grSU(2)$
global symmetry, which is the R-symmetry in $\superN=4$ supersymmetric Chern-Simons gauge theories.

\begin{figure}
\begin{center}
\includegraphics{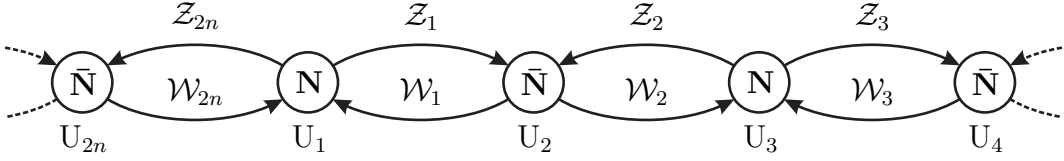}
\caption{\textbf{Non-chiral quiver.} The fields $\ZZ_\ell$ transform in $(\rep{N},\rep{\bar{N}})$ representations and the fields $\WW_\ell$ in $(\rep{\bar{N}},\rep{N})$ ones. The arrows indicate under which of the $\grU_\ell(N)$ the fields transform. For instance $\ZZ_2$ transforms under $(\grU_{3}(N),\grU_{2}(N))$. We close the chain by identifying $\grU_{2n+1}(N) \equiv \grU_{1}(N)$.}
\label{fig:quiverchain}
\end{center}
\end{figure}

To construct the gauge theory, which turns out to be a non-abelian generalization of the $\superN=4$ supersymmetric quiver gauge theory found in Sec.\ 3.2 of \cite{Hosomichi:2008jd}, we apply the well-known orbifold projection technique introduced in 
\cite{Douglas:1996sw}. The starting point is the gauge theory from \secref{sec:N-N} for gauge group $\grU(nN)\times \grU(nN)$. We rename the fields as $\ZZ\rightarrow\ZZZ$, $\WW\rightarrow\WWW$, $\VV\rightarrow\VVV$ and $\hat{\VV}\rightarrow\hat{\VVV}$ in order to have the original variables available for the fields after the orbifold projection. The $\Integers_n$ orbifold action is given by
\begin{align} \label{eqn:non-chiral-orbifold-action}
 \ZZZ^1     & \rightarrow e^{2\pi i/n} \Omega \ZZZ^1 \Omega^\dagger \; , &
 \WWW_1     & \rightarrow e^{-2\pi i/n} \Omega \WWW_1 \Omega^\dagger \; , &
 \VVV       & \rightarrow \Omega \VVV \Omega^\dagger \; , \\ \nn
 \ZZZ^2     & \rightarrow \Omega \ZZZ^2 \Omega^\dagger \; , &
 \WWW_2     & \rightarrow \Omega \WWW_2 \Omega^\dagger \; , &
 \hat{\VVV} & \rightarrow \Omega \hat \VVV \Omega^\dagger \; ,
\end{align}
where $\Omega=\diag(\unit_{N\times N}, e^{2\pi i/n} \unit_{N\times N}, \ldots, e^{2\pi i(n-1)/n} \unit_{N\times N})$. In the orbifold theory only those components of the superfields are retained which are invariant under \eqref{eqn:non-chiral-orbifold-action}. Explicitly these components are
\be
  \ZZZ^1 & = \matr{ccccc}{
  0          & \ZZ_1 &        &        & \\
             & 0     & \ZZ_3  &        & \\
             &       & 0      & \ddots & \\
             &       &        & 0      & \ZZ_{2n-3}\\
  \ZZ_{2n-1} &       &        &        & 0
  }
\comma
  \ZZZ^2 = \diag(\ZZ_{2n}, \ZZ_2, \ldots , \ZZ_{2n-2})
  \; ,
\ee
\be
  \WWW_1 & = \matr{ccccc}{
  0     &       &        &            & \WW_{2n-1} \\
  \WW_1 & 0     &        &            & \\
        & \WW_3 & 0      &            & \\
        &       & \ddots & 0          & \\
        &       &        & \WW_{2n-3} & 0
  }
\comma
  \WWW_2 = \diag(\WW_{2n}, \WW_2, \ldots , \WW_{2n-2})
  \; ,
\ee
and
\be
  \VVV = \diag(\VV_1, \VV_3, \ldots , \VV_{2n-1})
  \comma
  \hat{\VVV} = \diag(\VV_{2n}, \VV_2, \ldots , \VV_{2n-2})
  \; .
\ee
The projection has broken the $\grU(nN)\times\grU(nN)$ gauge symmetry down to the $\grU_1(N)\otimes\ldots\otimes\grU_{2n}(N)$ and the new field content is given by $\VV_\ell$, $\ZZ_\ell$, $\WW_\ell$, $\bar{\ZZ}_\ell$ and $\bar{\WW}_\ell$ for $\ell = 1,\ldots 2n$. These matter fields transform under bifundamental representations of various pairs of $\grU(N)$'s. Our labeling is such that the rows of $\ZZZ$ correspond to the gauge groups $\grU_1,\grU_3,\ldots,\grU_{2n-1}$ and the columns to $\grU_{2n},\grU_2,\grU_4,\ldots,\grU_{2n-2}$. For $\WWW$ rows and columns are interchanged. These transformations properties are depicted and further described in the quiver diagram in \figref{fig:quiverchain}.

The action of the orbifold theory is given by the following Chern-Simons and matter parts,
\be
  \Action_{\mathrm{CS}} \eq i K \int d^3x\,d^4\theta \int_0^1 dt\: \sum_{\ell=1}^{2n} (-1)^\ell 
  \tr \VV_\ell \bar{D}^\alpha \Bigbrk{ e^{t \VV_\ell} D_\alpha e^{-t \VV_\ell} } \; , \\
  \Action_{\mathrm{mat}} \eq \int d^3x\,d^4\theta\: \sum_{\ell=1}^{n} \tr \Bigsbrk{
   - \bar{\ZZ}_{2\ell-1} e^{-\VV_{2\ell-1}} \ZZ_{2\ell-1} e^{\VV_{2\ell  }} 
   - \bar{\WW}_{2\ell-1} e^{-\VV_{2\ell  }} \WW_{2\ell-1} e^{\VV_{2\ell-1}}
\nl[-1mm] \hspace{30mm}
   - \bar{\ZZ}_{2\ell  } e^{-\VV_{2\ell+1}} \ZZ_{2\ell  } e^{\VV_{2\ell  }}
   - \bar{\WW}_{2\ell  } e^{-\VV_{2\ell  }} \WW_{2\ell  } e^{\VV_{2\ell+1}}
   }
\ee
and the superpotential
\be
  \superW \eq \sum_{\ell=1}^{n} \Half \tr \Bigsbrk{ 
    \ZZ_{2\ell-1} \WW_{2\ell  } \ZZ_{2\ell  } \WW_{2\ell-1}
  - \ZZ_{2\ell  } \WW_{2\ell  } \ZZ_{2\ell+1} \WW_{2\ell+1}
  } \\
  \bar{\superW} \eq \sum_{\ell=1}^{n} \Half \tr \Bigsbrk{ 
    \bar{\ZZ}_{2\ell  } \bar{\WW}_{2\ell+1} \bar{\ZZ}_{2\ell+1} \bar{\WW}_{2\ell  }
  - \bar{\ZZ}_{2\ell-1} \bar{\WW}_{2\ell-1} \bar{\ZZ}_{2\ell  } \bar{\WW}_{2\ell  }
  } \; .
\ee
The terms in this superpotential correspond to all closed loops that follow four arrows and connect three sites in the quiver diagram in \figref{fig:quiverchain}. For $n=1$ we get back the original ABJM theory \cite{Aharony:2008ug} of \secref{sec:N-N}.

We note that the orbifold projection has partially broken the $\grSU(4)$ R-symmetry of the $n=1$ model. 
Since the fields do not carry any further index besides the gauge indices and the label $\ell$, only a 
$\grU(1)_R$ symmetry is manifest. However, similarly to the previous cases, we observe an R-symmetry 
enhancement to $\grSU(2)_o\times \grSU(2)_e$ which indicates that the theory possesses $\superN=4$ 
supersymmetry.\footnote{This agrees with the conclusion reached in section 3.2 of \cite{Hosomichi:2008jd} 
about the abelian, $N=1$, version of this gauge theory.}
The $\grSU(2)_o\times \grSU(2)_e$ symmetry can be made manifest in the potential by introducing doublets
\be
  Y^A_\ell = \{ Z_\ell, W^\dagger_\ell \}
  \comma
  Y^\dagger_{A,\ell} = \{ Z^\dagger_\ell, W_\ell \}
  \; ,
\ee
for each link $\ell$. Then the $\grSU(2)_o$ factor rotates the fields on the odd links, and the $\grSU(2)_e$ factor those on the even links. In order to illustrate this statement, we write down the bosonic potential:
\newcommand{\h}{\hphantom{+1}}
\begin{equation} \label{eqn:circularquiver-Vbos}
\begin{split}
   V^{\mathrm{bos}} = - \frac{L^2}{48} \sum_{\ell=1}^{n} \Big[ \hspace{8mm}
     & \tr Y^A_{2\ell\h} Y_{A,2\ell\h}^\dagger Y^B_{2\ell\h} Y_{B,2\ell\h}^\dagger Y^C_{2\ell\h} Y_{C,2\ell\h}^\dagger \\[-4mm]
+  3 & \tr Y^A_{2\ell\h} Y_{A,2\ell\h}^\dagger Y^B_{2\ell\h} Y_{B,2\ell\h}^\dagger Y^C_{2\ell+1} Y_{C,2\ell+1}^\dagger \\
+  3 & \tr Y^A_{2\ell\h} Y_{A,2\ell\h}^\dagger Y^B_{2\ell+1} Y_{B,2\ell+1}^\dagger Y^C_{2\ell+1} Y_{C,2\ell+1}^\dagger \\
+    & \tr Y^A_{2\ell+1} Y_{A,2\ell+1}^\dagger Y^B_{2\ell+1} Y_{B,2\ell+1}^\dagger Y^C_{2\ell+1} Y_{C,2\ell+1}^\dagger \\[3mm]
+    & \tr Y_{A,2\ell-1}^\dagger Y^A_{2\ell-1} Y_{B,2\ell-1}^\dagger Y^B_{2\ell-1} Y_{C,2\ell-1}^\dagger Y^C_{2\ell-1} \\
+  3 & \tr Y_{A,2\ell-1}^\dagger Y^A_{2\ell-1} Y_{B,2\ell-1}^\dagger Y^B_{2\ell-1} Y_{C,2\ell\h}^\dagger Y^C_{2\ell\h} \\
+  3 & \tr Y_{A,2\ell-1}^\dagger Y^A_{2\ell-1} Y_{B,2\ell\h}^\dagger Y^B_{2\ell\h} Y_{C,2\ell\h}^\dagger Y^C_{2\ell\h} \\
+    & \tr Y_{A,2\ell\h}^\dagger Y^A_{2\ell\h} Y_{B,2\ell\h}^\dagger Y^B_{2\ell\h} Y_{C,2\ell\h}^\dagger Y^C_{2\ell\h} \\[3mm]
+  4 & \tr Y^A_{2\ell-1} Y_{B,2\ell-1}^\dagger Y^C_{2\ell-1} Y_{A,2\ell-1}^\dagger Y^B_{2\ell-1} Y_{C,2\ell-1}^\dagger \\
+ 12 & \tr Y^A_{2\ell\h} Y_{B,2\ell\h}^\dagger Y^C_{2\ell+1} Y_{A,2\ell+2}^\dagger Y^B_{2\ell+2} Y_{C,2\ell+1}^\dagger \\
+ 12 & \tr Y^A_{2\ell+1} Y_{B,2\ell+1}^\dagger Y^C_{2\ell\h} Y_{A,2\ell-1}^\dagger Y^B_{2\ell-1} Y_{C,2\ell\h}^\dagger \\
+  4 & \tr Y^A_{2\ell\h} Y_{B,2\ell\h}^\dagger Y^C_{2\ell\h} Y_{A,2\ell\h}^\dagger Y^B_{2\ell\h} Y_{C,2\ell\h}^\dagger 
   \end{split}
\end{equation}
\begin{equation} \nn
\begin{split}
-  6 & \tr Y^A_{2\ell-1} Y_{B,2\ell-1}^\dagger Y^B_{2\ell-1} Y_{A,2\ell-1}^\dagger Y^C_{2\ell-1} Y_{C,2\ell-1}^\dagger \\
-  6 & \tr Y^A_{2\ell\h} Y_{B,2\ell\h}^\dagger Y^B_{2\ell\h} Y_{A,2\ell\h}^\dagger Y^C_{2\ell\h} Y_{C,2\ell\h}^\dagger \\
-  6 & \tr Y^A_{2\ell+1} Y_{B,2\ell+1}^\dagger Y^B_{2\ell+1} Y_{A,2\ell+1}^\dagger Y^C_{2\ell\h} Y_{C,2\ell\h}^\dagger \\
-  6 & \tr Y^A_{2\ell\h} Y_{B,2\ell\h}^\dagger Y^B_{2\ell\h} Y_{A,2\ell\h}^\dagger Y^C_{2\ell+1} Y_{C,2\ell+1}^\dagger \\
-  6 & \tr Y^A_{2\ell-1} Y_{B,2\ell\h}^\dagger Y^B_{2\ell\h} Y_{A,2\ell-1}^\dagger Y^C_{2\ell-1} Y_{C,2\ell-1}^\dagger \\
-  6 & \tr Y^A_{2\ell\h} Y_{B,2\ell-1}^\dagger Y^B_{2\ell-1} Y_{A,2\ell\h}^\dagger Y^C_{2\ell\h} Y_{C,2\ell\h}^\dagger \\
-  6 & \tr Y^A_{2\ell+1} Y_{B,2\ell+2}^\dagger Y^B_{2\ell+2} Y_{A,2\ell+1}^\dagger Y^C_{2\ell\h} Y_{C,2\ell\h}^\dagger \\[-1mm]
-  6 & \tr Y^A_{2\ell\h} Y_{B,2\ell-1}^\dagger Y^B_{2\ell-1} Y_{A,2\ell\h}^\dagger Y^C_{2\ell+1} Y_{C,2\ell+1}^\dagger
   \hspace{3mm} \Big] \; .
   \end{split}
\end{equation}
As a matter of fact, this potential is almost $\grSU(2)^{2n}$ invariant. Only the two terms with factor $12$ break this symmetry to $\grSU(2)_o\times \grSU(2)_e$.

\paragraph{Note added.} After the original version of this paper appeared, two papers 
[\citen{Imamura:2008nn},\citen{Terashima:2008ba}]
analyzed the moduli space of this non-chiral $\grU(N)^{2n}$ quiver gauge theory for $N=1$. These papers demonstrate that one of the $\grU(1)$ gauge symmetries, which corresponds to the combination of the gauge potentials $\sum_{\ell=1}^{2n} (-1)^{\ell} A_\ell$, is broken to a discrete subgroup. Assuming their choice of quantization condition is correct, this implies that for the non-chiral quiver chain the moduli space is $\Complex^4/(\Integers_n\times \Integers_{kn})$, i.e.\ $\tilde{k}=kn$. It is thus natural to conjecture that the gauge theory describes $N$ coincident M2-branes on this orbifold.

\section{Chiral orbifold gauge theories}
\label{sec:chiralquiver}

In this section we consider a different orbifold projection of the non-chiral ABJM theory, which produces a gauge theory whose matter fields do not form pairs in mutually conjugate representations. The $\Integers_l$ action is now given by
\be
y^1\rightarrow e^{2\pi i/l} y^1 \; , \quad
y^2\rightarrow e^{-2\pi i/l} y^2 \; , \quad
y^3\rightarrow  y^3 \; , \quad
y^4\rightarrow  y^4 \; .
\ee
This rotates the spinors of $\grSO(8)$ by the phase $e^{2\pi i(s_1-s_2)/l}$. Thus, the spinors left invariant by the combined $\Integers_k$ and $\Integers_l$ actions have $2s_1+s_3+s_4=0$. There are 2 such spinors corresponding to $\superN=2$ supersymmetry. The orbifold action also preserves $\grSU(2)$ global symmetry which commutes with the $\grU(1)_R$ symmetry (in the special case $l=2$ this global symmetry is actually enhanced to $\grSU(2)\times \grSU(2)$).

\begin{figure}
\begin{center}
\includegraphics{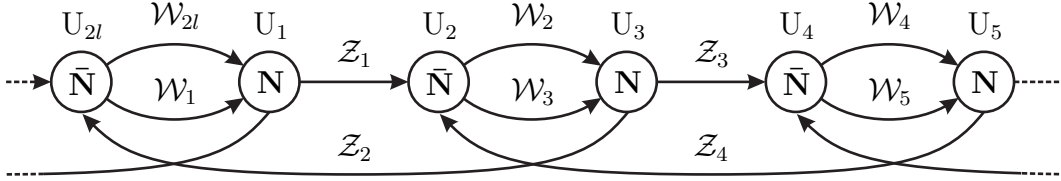}
\caption{\textbf{Chiral quiver.} The characteristic property of the chiral quiver is that no nodes are connected by chiral fields with mutually conjugate representation (no anti-parallel arrows).}
\label{fig:chiralquiver}
\end{center}
\end{figure}

On the gauge theory side, we start with the $\grU(lN)\times \grU(lN)$ gauge theory with fields $\ZZZ$, $\WWW$, $\VVV$, $\hat{\VVV}$, and retain the superfields invariant under the $\Integers_l$ action
\begin{align}
 \ZZZ^1     & \rightarrow e^{2\pi i/l} \Omega \ZZZ^1 \Omega^\dagger \; &
 \WWW_1     & \rightarrow \Omega \WWW_1 \Omega^\dagger \; , &
 \VVV       & \rightarrow \Omega \VVV \Omega^\dagger \; , \\
 \ZZZ^2     & \rightarrow e^{-2\pi i/l} \Omega \ZZZ^2 \Omega^\dagger \; , &
 \WWW_2     & \rightarrow \Omega \WWW_2 \Omega^\dagger \; , &
 \hat{\VVV} & \rightarrow \Omega \hat \VVV \Omega^\dagger \; ,
\end{align}
where $\Omega=\diag(\unit_{N\times N}, \, e^{2\pi i/l} \unit_{N\times N} \, , \ldots, \, e^{2\pi i(l-1)/l} \unit_{N\times N})$. This projection breaks the $\grU(lN) \times\grU(lN)$ gauge symmetry down to the $(\grU(N)\times \grU(N))^l$, and the resulting chiral field content is summarized in the quiver diagram in \figref{fig:chiralquiver}. The global $\grSU(2)$ symmetry acts on the pairs of fields having parallel arrows. Now there is no possibility of a non-trivial symmetry restoration since on each link there are no pairs of chiral superfields in mutually conjugate representations of the gauge group. Thus, there is no extended supersymmetry, and we are dealing with an $\superN=2$ gauge theory that just happens to possess additional global $\grSU(2)$ symmetry. The symmetries match those of a $\Integers_l\times \Integers_{\tilde{k}}$ orbifold where the $\grSU(2)$ symmetry corresponds to rotations of $y^3$ and $y^4$.

The chiral $\Integers_2$ projection corresponding to $l=2$ was originally considered in $3+1$ dimensions to construct the theory on D3-branes at the tip of the cone over $T^{11}/\Integers_2$ where the $\Integers_2$ acts freely \cite{Morrison:1998cs}. In this case the orbifold projection does not break the $\grSU(2)\times \grSU(2)$ symmetry of the quartic superpotential. For the generalization to $l>2$ in the context of conifold theory, which preserves only one global $\grSU(2)$, see for example \cite{Benvenuti:2004dy}.

\section{$\superN=2$ superconformal theory with $\grSU(3)$ symmetry and RG flow}
\label{sec:supersu3}

Let us consider the $\grU(N)\times \grU(N)$ gauge theories in the special cases $k=1$ or $k=2$, where
they are expected to possess a global $\grSU(4)$ non-R symmetry. 
We can then add a relevant superpotential deformation that breaks it to $\grSU(3)$, and
this RG flow could take the theory to a new $\superN=2$ superconformal theory with $\grSU(3)$ symmetry. 
An analogous construction in the 3+1 dimensional $\superN=4$ SYM theory is to add a quadratic term 
in one of the superfields, which creates an RG flow leading to a $\grU(1)_R\times \grSU(2)$
invariant superconformal theory with a quartic superpotential \cite{Freedman:1999gp}. 
Ideas similar to this were explored also in the $2+1$ dimensional case \cite{Johnson:2001ze} where
a quadratic term breaks the $\grSU(4)$ global symmetry to $\grSU(3)$; 
we will make them more concrete here.
 
A subtlety of the ABJM theories with $N>2$ is that only the $\grSU(2)\times \grSU(2)$ subgroup of the global $\grSU(4)_R$ is manifest in the superpotential \cite{Aharony:2008ug}. We will thus consider the $N=2$ case, closely related to the BLG theory, where a global $\grSU(4)$ is manifest in the superpotential \eqref{eqn:superpot4}. A simple quadratic deformation that preserves the gauge symmetry and an $\grSU(3)$ gives the superpotential
\be \label{massdef}
  \superW  = -\frac{1}{8\cdot 4!} \; \levi_{ABCD} \levi^{abcd} \ZZ^A_a \ZZ^B_b \ZZ^C_c \ZZ^D_d  
        + m (\ZZ^4_a)^2 \; .
\ee
This relevant deformation causes RG flow that takes the theory to an $\superN=2$ superconformal theory whose effective superpotential is found by integrating out $\ZZ^4_a$:
\be
  \superW_{\rm eff} \sim \; (\levi_{ABC} \levi^{abcd} \ZZ^A_a \ZZ^B_b \ZZ^C_c )^2
\; .
\ee
This sextic superpotential is marginal if we assign R-charge $1/3$ to the remaining superfields $\ZZ^A_a$, $A=1,2,3$. Since $\theta$ has R-charge $1$, the fermionic superpartner has R-charge $-2/3$. This means that the exact scaling dimension of the bosonic fields $Z^A_a$ is $1/3$, and of their fermionic superpartners is $5/6$. In addition to the $\grU(1)_R$ symmetry, the superpotential is invariant under a global $\grSU(3)$ symmetry that acts on the index $A$. 

Thus, we have found $\superN=2$ superconformal Chern-Simons theories with global $\grSU(3)$ symmetry, and we need to search for their M-theory duals. Remarkably, N. Warner \cite{Warner:1983vz} has found an AdS$_4$ extremum of the $\superN=8$ gauged supergravity \cite{deWit:1982ig} with exactly the same symmetries as our gauge theory; namely, $\grU(1)_R\times \grSU(3)$. Its uplifting to 11 dimensions produces a warped product of $AdS_4$ and a ``stretched and squashed'' 7-sphere \cite{Corrado:2001nv,Ahn:2002eh}, which contains a $CP^2$ giving rise to the $\grSU(3)$ symmetry. We can plausibly conjecture that the $\grU(2)\times \grU(2)$ Chern-Simons gauge theory with level $k=1$ and the sextic superpotential $\superW_{\mathrm{eff}}$ is dual to such a background supported by two units of $G_4$ flux (but the supergravity approximation applies only in the limit of large flux). The theory at level $k=2$ is then dual to a $\Integers_2$ orbifold of the background in \cite{Corrado:2001nv,Ahn:2002eh}.

In fact, the full holographic RG flow from the $\grSO(8)_R$ symmetric AdS$_4$ extremum in the UV to the $\grU(1)_R\times \grSU(3)$ symmetric AdS$_4$ extremum in the IR was constructed in \cite{Ahn:2000aq}, and its uplifting to 11 dimensions in \cite{Corrado:2001nv}. It was shown that the relevant operator giving rise to the RG flow in the dual gauge theory has dimension 2 \cite{Ahn:2000aq}. This precisely agrees with the dimension of the fermion bilinear we have added to the potential. Further studies of the holographic RG flow \cite{Johnson:2001ze} showed that in the IR theory there are chiral superfields of R-charge $1/3$, consistent with our claim. We may therefore conjecture that this RG flow is encoded in the superpotential \eqref{massdef}. 

In order to check the AdS$_4$/CFT$_3$ correspondence we are proposing, we should match the R-charges and 
dimensions of the gauge invariant operators. Luckily, in gauged supergravity the spectrum of 
perturbations was analyzed long ago \cite{Nicolai:1985hs}, and we will use these results. 
Perhaps the simplest chiral operators we can write down transform in the $\rep{6}$ of $\grSU(3)$:
\be \label{sextet}
\ZZ^{(A}_a \ZZ^{B)}_a
\ee
This multiplet of operators consists of a scalar field of $R$-charge $2/3$ and dimension $2/3$;
a spin $1/2$ fermion of $R$-charge $-1/3$ and dimension $7/6$; and a pseudoscalar 
of $R$-charge $-4/3$ and dimension $5/3$. In \cite{Nicolai:1985hs} the fields with such quantum
numbers can be found in Table 2 corresponding to a massive hypermultiplet. 
It is further stated that there is a sextet with R-charge $y=2/3$, 
which agrees with \eqref{sextet}. 
The corresponding operator dimension is \cite{Nicolai:1985hs} 
$E_0=\lambda^{-1} |y|$, and the standard relation
between dimension and R-charge in $2+1$ dimensions requires $\lambda=1$ (this differs from the
assignment $\lambda=1/2$ made in \cite{Nicolai:1985hs}). In fact,
using $\lambda=1$ and $y=\pm 2/3$ in Table 2 of \cite{Nicolai:1985hs} we match the R-charges
and dimensions of the operators contained in the supermultiplet \eqref{sextet}, as well as
in the corresponding anti-chiral supermultiplet. 

Clearly, it is necessary to subject these ideas to further tests.
One obvious problem is to construct explicitly the $\grU(1)_R\times \grSU(3)$ symmetric 
$\grU(N)\times \grU(N)$ gauge theory for $N>2$, which we conjecture to be dual to the background 
of \cite{Corrado:2001nv} with $N$ units of flux.

\section{Discussion}

In this letter we made first steps towards generalizing the BLG \cite{Bagger:2006sk, Bagger:2007jr,Bagger:2007vi,Gustavsson:2007vu} and ABJM \cite{Aharony:2008ug} superconformal Chern-Simons gauge theories. We wrote down the superspace formulation for the $\grU(N)\times \grU(N)$ Chern-Simons theories with bifundamental matter and a quartic superpotential of \cite{Klebanov:1998hh}, which at level $k$ describe $N$ M2-branes at a $\Integers_k$ orbifold \cite{Aharony:2008ug}. We also wrote down new theories describing $N$ M2-branes at certain singularities $\Complex^4/(\Integers_n\times \Integers_{kn})$. In \secref{sec:quivercircle} we presented a family of non-chiral quiver gauge theories which have $\superN=4$ supersymmetry, and in \secref{sec:chiralquiver} a family of chiral quiver gauge theories 
possessing $\superN=2$ supersymmetry. Finally, we conjectured that $k=1$ Chern-Simons gauge theories with $\grSU(3)$ invariant relevant superpotential deformation are dual to the holographic RG flows constructed in \cite{Ahn:2000aq,Corrado:2001nv}.

Clearly, there are many possible further generalizations of this work.
It would be desirable to understand systematically the gauge theories describing M2-branes at arbitrary
orbifold singularities. 
Consider, for example, the orbifolds $\Complex^4/\Integers_k$ where the $\grSU(3)$ symmetric action on the
4 complex coordinates is $\diag(e^{2\pi i/k}, e^{2\pi i/k}, e^{2\pi i/k}, e^{-6\pi i/k})$.
On spinors this translates into multiplication by $e^{2\pi i (s_1+ s_2 + s_3 - 3 s_4)/k }$,
and it is easy to see that the orbifold preserves $\superN=2$ supersymmetry.
It seems difficult, however, to write down the dual gauge theory that has manifest
global $\grSU(3)$ symmetry. To see the full $\grSU(3)$ one may need to invoke the 't Hooft operators used in \cite{Aharony:2008ug}.

It would also be interesting to study resolution of orbifolds. In the $3+1$ dimensional case
new theories may be obtained through turning on the Fayet-Iliopoulos terms which correspond to partial resolutions
of orbifolds \cite{Morrison:1998cs}. Perhaps a similar approach can be undertaken also in $2+1$ dimensions
to produce theories of M2-branes at more general conical singularities.

\section*{Acknowledgments}
We are very grateful to Juan Maldacena for explaining his results prior to publication and for other 
useful discussions. I.K. also thanks Chris Herzog, Arvind Murugan and John Schwarz for discussions, 
and the Aspen Center for Physics for hospitality. This research is supported in part by the National Science Foundation Grant No.~PHY-0756966 and in part by a Marie Curie Outgoing International Fellowship, contract No. MOIF-CT-2006-040369, within the 6th European Community Framework Programme.

\appendix

\section{Notation}
\label{sec:notations}

The world-volume metric is $g^{\mu\nu} = \diag(-1,+1,+1)$ with index range $\mu=0,1,2$. We use Dirac matrices $(\gamma^\mu)_\alpha{}^\beta = (i \sigma^2, \sigma^1, \sigma^3)$ satisfying $\gamma^\mu \gamma^\nu = g^{\mu\nu} + \levi^{\mu\nu\rho} \gamma_\rho$. The fermionic coordinate of superspace is a complex two-component spinor $\theta$. Indices are raised, $\theta^\alpha = \levi^{\alpha\beta} \theta_\beta$, and lowered, $\theta_\alpha = \levi_{\alpha\beta} \theta^\beta$, with $\levi^{12} = -\levi_{12} = 1$. Note that lowering the spinor indices of the Dirac matrices makes them symmetric $\gamma^\mu_{\alpha\beta} = (-\unit, -\sigma^3, \sigma^1)$. In products like $\theta^\alpha \theta_\alpha \equiv \theta^2$, $\theta^\alpha \bar{\theta}_\alpha \equiv \theta \bar{\theta}$ etc and $\theta^\alpha \gamma^\mu_{\alpha\beta} \bar{\theta}^\beta \equiv \theta \gamma^\mu \bar{\theta}$ we suppress the indices. We have
\be
  \theta_\alpha \theta_\beta = \half \levi_{\alpha\beta} \theta^2
  \comma
  \theta^\alpha \theta^\beta = \half \levi^{\alpha\beta} \theta^2
\ee
and likewise for $\bar{\theta}$ and derivatives. The Fierz identities are\footnote{Here and everywhere we use symmetrization and anti-symmetrization with weight one $X_{[a} Y_{b]} = \half \bigbrk{ X_a Y_b - X_b Y_a }$, $X_{(a} Y_{b)} = \half \bigbrk{ X_a Y_b + X_b Y_a }$.}
\be
  (\psi_1 \psi_2) (\psi_3 \psi_4) \eq
  - \half (\psi_1 \psi_4) (\psi_3 \psi_2)
  - \half (\psi_1 \gamma^\mu \psi_4) (\psi_3 \gamma_\mu \psi_2) \; , \\
\nn
  (\psi_1 \psi_2) (\psi_3 \gamma^\mu \psi_4) \eq
  - \half (\psi_1 \gamma^\mu \psi_4) (\psi_3 \psi_2)
  - \half (\psi_1 \psi_4) (\psi_3 \gamma^\mu \psi_2)
  + \half \levi^{\mu\nu\rho} (\psi_1 \gamma_\nu \psi_4) (\psi_3 \gamma_\rho \psi_2) \; , \\
\nn
  (\psi_1 \gamma^\mu \psi_2) (\psi_3 \gamma^\nu \psi_4) \eq
  - \half g^{\mu\nu} (\psi_1 \psi_4) (\psi_3 \psi_2)
  + \half g^{\mu\nu} (\psi_1 \gamma^\rho \psi_4) (\psi_3 \gamma_\rho \psi_2)
  - (\psi_1 \gamma^{(\mu} \psi_4) (\psi_3 \gamma^{\nu)} \psi_2) \nl \nn
  + \half \levi^{\mu\nu\rho} \bigsbrk{ (\psi_1 \gamma_\rho \psi_4) (\psi_3 \psi_2)
                                     - (\psi_1 \psi_4) (\psi_3 \gamma_\rho \psi_2) } \; ,
\ee
which imply in particular
\be
  (\theta\bar{\theta})^2
  = -\half \theta^2 \bar{\theta}^2
  \comma
  (\theta\bar{\theta})(\theta\gamma^\nu\bar{\theta})
  = 0
  \comma
  (\theta\gamma^\mu\bar{\theta})(\theta\gamma^\nu\bar{\theta})
  = \half g^{\mu\nu} \theta^2 \bar{\theta}^2
  \; .
\ee
Supercovariant derivatives and susy generators are
\begin{align}
  D_\alpha & = \partial_\alpha + i (\gamma^\mu \bar{\theta})_\alpha \partial_\mu \; , &
  Q_\alpha & = \partial_\alpha - i (\gamma^\mu \bar{\theta})_\alpha \partial_\mu \; , \\
  \bar{D}_\alpha & = -\bar{\partial}_\alpha - i (\theta\gamma^\mu)_\alpha \partial_\mu \; , &
  \bar{Q}_\alpha & = -\bar{\partial}_\alpha + i (\theta\gamma^\mu)_\alpha \partial_\mu \; ,
\end{align}
with the only non-trivial anti-commutators
\be
  \acomm{D_\alpha}{\bar{D}_\beta} = -2i \gamma^\mu_{\alpha\beta} \partial_\mu
  \comma
  \acomm{Q_\alpha}{\bar{Q}_\beta} = 2i \gamma^\mu_{\alpha\beta} \partial_\mu
  \; .
\ee
We use the following conventions for integration
\be
  d^2\theta \equiv -\quarter d\theta^\alpha d\theta_\alpha 
  \comma
  d^2\bar{\theta} \equiv - \quarter d\bar{\theta}^\alpha d\bar{\theta}_\alpha 
  \comma
  d^4\theta \equiv d^2\theta \, d^2\bar{\theta}
  \; ,
\ee
such that
\be
  \int d^2\theta \, \theta^2 = 1
  \comma
  \int d^2\bar{\theta} \, \bar{\theta}^2 = 1
  \comma
  \int d^4\theta \, \theta^2 \bar{\theta}^2 = 1
  \; .
\ee
It is useful to note that up to a total derivative
\be
  \int d^4\theta \, \ldots = \frac{1}{16} \eval{\bigbrk{D^2 \bar{D}^2 \ldots}}_{\theta=\bar{\theta}=0} \; .
\ee

The components of a chiral and an anti-chiral superfield, $\ZZ(x_L,\theta)$ and $\bar{\ZZ}(x_R,\bar{\theta})$, are a complex boson $\phi$, a complex two-component fermion $\psi$ and a complex auxiliary scalar $F$. Their component expansions are given by
\be
  \ZZ = \phi(x_L) + \sqrt{2} \, \theta\psi(x_L) + \theta^2 \, F(x_L)
\comma
  \bar{\ZZ} = \bar{\phi}(x_R) - \sqrt{2} \, \bar{\theta}\bar{\psi}(x_R) - \bar{\theta}^2 \, \bar{F}(x_R)
\ee
where
\be
  x_L^\mu = x^\mu + i \theta \gamma^\mu \bar{\theta}
  \comma
  x_R^\mu = x^\mu - i \theta \gamma^\mu \bar{\theta} \; .
\ee
The components of the gauge superfield $\VV(x,\theta,\bar{\theta})$ in Wess-Zumino gauge are the gauge field $A_\mu$, a complex two-component fermion $\chi_\alpha$, a real scalar $\sigma$ and an auxiliary scalar $\auxD$, such that
\be
  \VV = 2i \, \theta \bar{\theta} \, \sigma(x)
    + 2 \, \theta\gamma^\mu\bar{\theta} \, A_\mu(x)
    + \sqrt{2} i \, \theta^2 \, \bar{\theta}\bar{\chi}(x) 
    - \sqrt{2} i \, \bar{\theta}^2 \, \theta \chi(x)
    + \theta^2 \, \bar{\theta}^2 \, \auxD(x)
    \; .
\ee

\bigskip

We use the $N\times N$ hermitian matrix generators $T^n$ ($n=0,\ldots, N^2-1$) and $t^n$ ($n=1,\ldots, N^2-1$) for $\grU(N)$ and $\grSU(N)$ respectively. We have $T^n = (T^0, t^n)$ with $T^0 = \unit/\sqrt{N}$. The generators are normalized as $\tr T^n T^m = \delta^{nm}$. Completeness implies $\tr A T^n \tr B T^n = \tr A B$, $\tr A T^n B T^n = \tr A \tr B$ for $\grU(N)$ and 
$\tr A t^n \tr B t^n = \tr A B - \tfrac{1}{N} \tr A \tr B$, $\tr A t^n B t^n = \tr A \tr B - \tfrac{1}{N} \tr A B$ for $\grSU(N)$.

\bibliographystyle{nb}
\bibliography{M2}

\end{document}